\documentclass[prl,twocolumn,amsmath,amssymb,superscriptaddress]{revtex4-2}

\usepackage{graphicx}
\usepackage{dcolumn}
\usepackage{bm}
\usepackage{ textcomp }
\usepackage{color}
\usepackage{amsmath}
\usepackage{amssymb}
\usepackage{xcolor}
\usepackage{easyReview}
\usepackage[colorlinks = true,
            linkcolor = blue,
            urlcolor  = red,
            citecolor = blue,
            anchorcolor = blue]{hyperref}
\usepackage{mathdots}
\usepackage{float}
\usepackage{lineno}
\usepackage{todonotes}
\usepackage{array}
\usepackage{xcolor,colortbl}
\definecolor{LightBlue}{rgb}{0.8,0.8,0.8}
\newcommand{\pdag}{{\phantom\dagger}}

\usepackage[normalem]{ulem}

\begin{document}
\title{Stripe correlations in the two-dimensional Hubbard-Holstein model}
\date{\today}

\author{Seher Karakuzu}
\affiliation{Computational Sciences and Engineering Division, Oak Ridge National Laboratory, Oak Ridge, TN 37831-6164, USA\looseness=-1}
\affiliation{Center for Computational Quantum Physics, Flatiron Institute, 162 5th Avenue, New York, NY 10010, USA\looseness=-1}

\author{Andy Tanjaroon Ly}
\affiliation{Department of Physics and Astronomy, The University of Tennessee, Knoxville, TN 37966, USA}

\author{Peizhi Mai}
\affiliation{Computational Sciences and Engineering Division, Oak Ridge National Laboratory, Oak Ridge, TN 37831-6164, USA\looseness=-1}
\affiliation{Department of Physics and Institute of Condensed Matter Theory, University of Illinois at Urbana-Champaign, Urbana, IL 61801, USA}

\author{James Neuhaus}
\affiliation{Department of Physics and Astronomy, The University of Tennessee, Knoxville, TN 37966, USA}

\author{Thomas A. Maier}
\affiliation{Computational Sciences and Engineering Division, Oak Ridge National Laboratory, Oak Ridge, TN 37831-6164, USA\looseness=-1}

\author{Steven Johnston}
\affiliation{Department of Physics and Astronomy, The University of Tennessee, Knoxville, TN 37966, USA}
\affiliation{Institute for Advanced Materials and Manufacturing, University of Tennessee, Knoxville, TN 37996, USA\looseness=-1}

\begin{abstract}
Several state-of-the-art numerical methods have observed static or fluctuating spin and charge stripes in doped two-dimensional Hubbard models, suggesting that these orders play a significant role in shaping the cuprate phase diagram. Many experiments, however, also indicate that the cuprates have strong electron-phonon ($e$-ph) coupling, and it is unclear how this interaction influences stripe correlations. We study static and fluctuating stripe orders in the doped singleband Hubbard-Holstein model using zero temperature variational Monte Carlo and finite temperature determinant quantum Monte Carlo. We find that the lattice couples more strongly with the charge component of the stripes, leading to an enhancement or suppression of stripe correlations, depending on model parameters like the next-nearest-neighbor hopping $t^\prime$ or phonon energy $\Omega$. Our results help elucidate how the $e$-ph interaction can tip the delicate balance between stripe and superconducting correlations in the Hubbard-Holstein model with implications for our understanding of the high-$T_\mathrm{c}$ cuprates. 
\end{abstract}
\maketitle

\noindent\textbf{Introduction} --- 
The prevailing view of the high-temperature (high-$T_\mathrm{c}$) cuprate superconductors is that they are governed by intertwined orders~\cite{FradkinRMP2015}. In this scenario, different coupled spin and charge orders (i.e., stripes) and their fluctuations compete/cooperate with unconventional superconductivity giving rise to a complex phase diagram~\cite{FradkinRMP2015, KeimerReview, TranquadaReview2021}. This behavior is observed not only experimentally \cite{TranquadaReview2021,TranquadaNature1995,PhysRevLett.85.4590,PhysRevB.70.104517,PhysRevB.78.174529,AbbamonteNaturePhys2005} but also in single- and multi-band Hubbard models \cite{ZaanenPRB1989,WhitePRL1989a, WhitePRL1989b, PhysRevB.39.9749,HusseinPRB2019,Zheng1155, PhysRevX.10.031016, MIYAZAKI201430, PhysRevB.97.045138, CorbozPRL2014, HuangScience2017,HuangQuantMat2018,JiangScience2019, PhysRevResearch.2.033073, PhysRevX.10.031016,sorella2021phase,HuangPreprint, mai2021stripes, XiaoPreprint}. For example, non-perturbative numerical methods that access zero-temperature properties frequently identify several nearly degenerate stripe and $d$-wave superconducting states for model parameters that are relevant to the cuprates \cite{MIYAZAKI201430, PhysRevB.97.045138, Zheng1155, PhysRevResearch.2.033073, JiangScience2019, PhysRevResearch.2.033073, sorella2021phase}. The state that ultimately wins out as the ground state, however, is sensitive to subtle factors like the value of the next-nearest-neighbor hopping $t^\prime$~\cite{JiangScience2019, PhysRevResearch.2.033073}. These results have cast doubt on the notion that the Hubbard model has a superconducting ground state for parameter regimes relevant to the cuprates~\cite{PhysRevX.10.031016, JiangScience2019, JiangPNAS2021}.

For finite temperature, quantum Monte Carlo (QMC) methods \cite{HuangScience2017, HuangQuantMat2018} find fluctuating spin stripes at high temperatures $T \sim 0.22t$, where $t$ is the nearest neighbor hopping integral. More recently, weaker charge stripe fluctuations have been reported in DCA simulations on large extended clusters~\cite{mai2021stripes}, where the charge correlations were observed to develop after the spin correlations as the temperature is lowered in the hole-doped system. This hierarchy of the spin and charge correlations appears to be a general property of the singleband Hubbard model at finite temperature; it has been observed in subsequent constrained-path auxiliary-field QMC \cite{XiaoPreprint} and DQMC \cite{HuangPreprint} simulations. Curiously, the behavior in real cuprates is reversed, where charge modulations tend to develop before spin modulations~\cite{TranquadaReview2021}. 

While it is clear that strong electron correlations dominate cuprate physics, there is also a growing body of evidence that electron-phonon ($e$-ph) interactions are also relevant \cite{LanzaraNature2001, PhysRevLett.93.117004, ShenPRL2004, LeeNature2006, LeePRB2007, JohnstonPRL2012, RossiPRL2019, ChenScience2021}. It is, therefore, essential to study the influence of $e$-ph interactions on stripe order, particularly in light of their apparent sensitivity to perturbing interactions. The singleband Hubbard-Holstein model is the minimal model describing correlated electrons coupled to the lattice. This model has been widely studied at half-filling, where competition between antiferromagnetic Mott and ${\bf Q} = (\pi/a,\pi/a)$ charge-density-wave (CDW) insulating phases is commonly observed~\cite{BauerPRB2010, NowadnickPRL2012, JohnstonPRB2013, MendlPRB2017, KarakuzuPRB2017, OhgoePRL2017, PhysRevB.98.085405, Costa2020}. Away from half-filling, there are suggestions that the $e$-ph interaction can enhance $d$-wave pairing correlations for some parameter regimes \cite{HuangPRB2003, PhysRevB.75.014503, MendlPRB2017}. However, we know comparatively little about how the $e$-ph interaction might influence stripe correlations and their competition with superconductivity. 

We present a study combining variational Monte Carlo (VMC) and determinant quantum Monte Carlo (DQMC) to examine static and fluctuating stripe correlations in the doped Hubbard-Holstein model. We find that the coupling to the lattice can enhance the charge component of the stripes while also suppressing their spin component, depending on the value of the specific model parameters. Our results show that the $e$-ph coupling can alter the balance between the stripe and superconducting correlations and suggest a potential solution to how charge-stripes might appear before spin-stripes in a real material. \\

\begin{figure*}[ht]
    \centering
    \includegraphics[width=\textwidth]{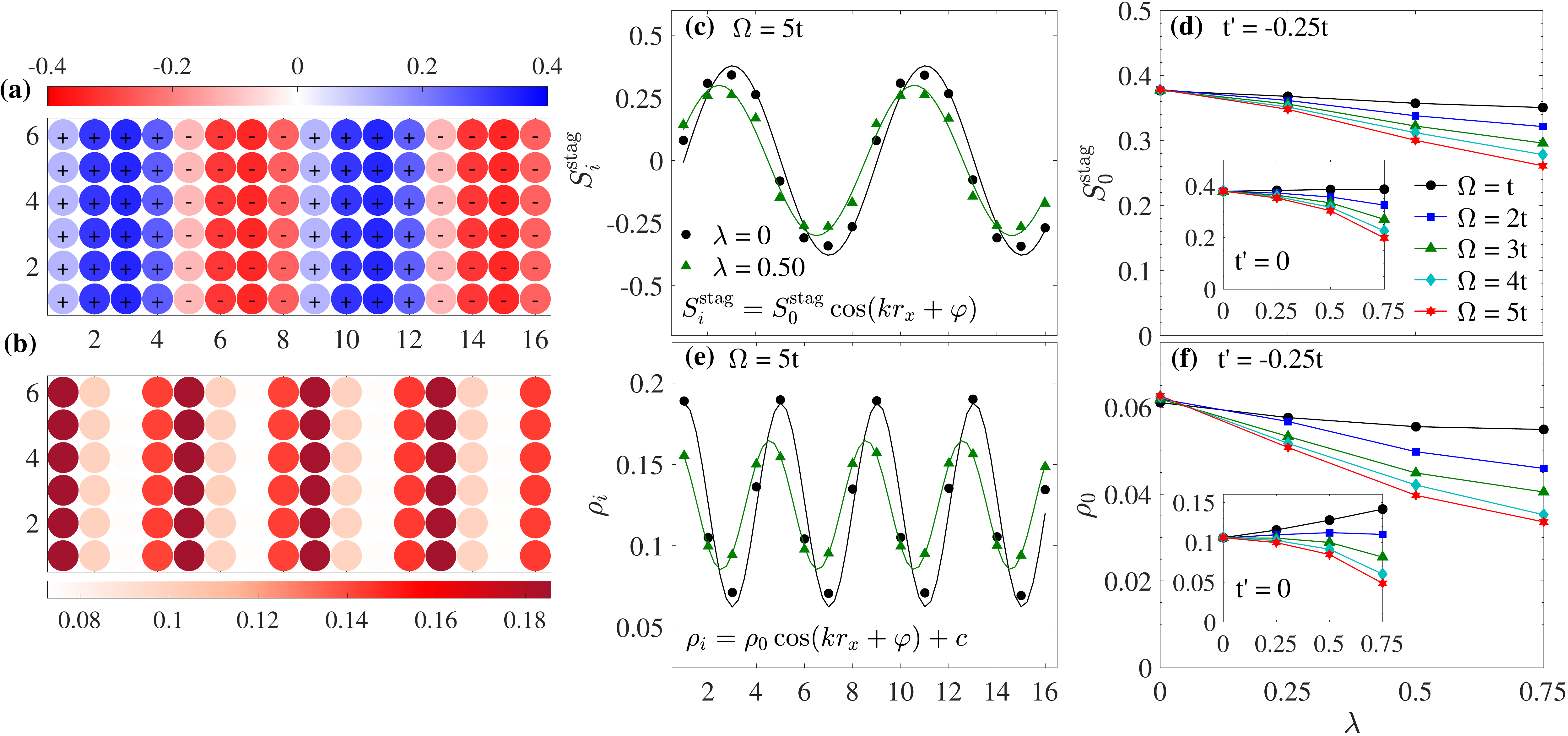}
    \caption{Variational Monte Carlo (VMC) results for static stripe order in the two-dimensional Hubbard-Holstein model. (a) The expectation value of the staggered local spin operator in real space $S_i^\mathrm{stag} = (-1)^{i_x+i_y}\langle \hat{S}_i^z\rangle$, where $S_i^z = \frac{1}{2}(\hat{n}_{i,\uparrow}-\hat{n}_{i,\downarrow})$ is the $z$-component of the local spin operator. Results are shown here for $U = 8t$, $t^\prime = -0.25t$, $\langle n \rangle = 0.875$, and $\lambda = 0$. (b) The expectation value of the local density operator $\rho_i = \langle \sum_\sigma \hat{n}_{i,\sigma} \rangle$ for the same case. Panels (c) and (e) show $S_i^\mathrm{stag}$ and $\rho_i$, respectively, along the line $(i_x,0)$ for representative values of $\lambda=0,~0.5$ and $\Omega = 5t$. The data points are the VMC values of the local quantity and the solid lines are sinusoidal fits to the VMC data. Panels (d) and (f) show the fitted amplitudes of the spin and charge modulations, respectively, as a function of $\lambda$ for $t^\prime = -0.25t$ and various $\Omega$. The insets of these panels show corresponding data for $t^\prime = 0$.}
    \label{fig:Stripes_VMC}
\end{figure*}

\noindent\textbf{Model} --- We study the singleband Hubbard-Holstein model, defined on a 
two-dimensional (2D) square lattice. Its Hamiltonian is 
$H=H_\mathrm{el}+H_\mathrm{ph}+H_{e-\mathrm{ph}}$, where 
\begin{equation*}
H_\mathrm{el} = -\mu\sum_{i}\hat{n}_{i} -\sum_{i,j,
\sigma} t^{\phantom\dagger}_{ij}c^\dagger_{i,\sigma}c^\pdag_{j,\sigma} + U\sum_{i}\hat{n}_{i,\uparrow}\hat{n}_{i,\downarrow}, 
\end{equation*}
describes the electronic subsystem, $H_\mathrm{ph} = \Omega\sum_{i} (b^\dagger_i b^\pdag_i+\frac{1}{2})$ describes the phononic subsystem, and $H_{e-\mathrm{ph}} = \sum_{i} g\hat{n}_i(b^\dagger_i + b_i^\pdag)$ describes their coupling. Here, $c^\dagger_{i,\sigma}$ ($c^\pdag_{i,\sigma}$) creates (annihilates) a spin-$\sigma$ ($=\uparrow,\downarrow$) electron on site $i$, 
$b^\dagger_{i}$ ($b^\pdag_{i}$) creates (annihilates) a dispersionless optical phonon at lattice site $i$ with energy $\Omega$,  $\hat{n}^\pdag_{i,\sigma}=c^\dagger_{i,\sigma}c^\pdag_{i,\sigma}$ and $\hat{n}_i =  \sum_\sigma \hat{n}^\pdag_{i,\sigma}$, $t_{ij}$ is the hopping integral between sites $i$ and $j$, ${\bf r}_i = a(i_x,i_y)$ ($i_{x(y)}\in \mathbb{Z}$) is a lattice vector, $\mu$ is the chemical potential, $U$ is the Hubbard repulsion, and $g$ is the $e$-ph coupling strength. Throughout, we set $t_{ij} = t$ for nearest neighbors, $t_{ij} = t^\prime$ for next-nearest neighbors, and $t_{ij} = 0$ otherwise, and vary $\Omega$ and $g$ over a range of values. Finally, we set $t = M = a= 1$ and adopt a standard parameterization of the dimensionless $e$-ph coupling with $\lambda = \frac{2g^2}{W\Omega} \approx \frac{g^2}{4t\Omega}$, where $W\approx 8t$ is the electronic bandwidth.  \\

\noindent\textbf{Methods} --- We study the model using VMC and DQMC, two nonperturbative numerical methods capable of treating both the $e$-$e$ and $e$-ph interactions on an equal footing. 
VMC is a zero-temperature method that uses Markov chain Monte Carlo to optimize a variational estimate for the system's ground state wave function. Here, we use the method as described in Ref.~\cite{KarakuzuPRB2017}, applied to rectangular $N = 16\times 6$ clusters with periodic boundary conditions (PBC). 
{We have considered two different variational wave functions which we label as uniform and stripe solutions. Our uniform wave function includes a $(\pi,\pi)$ antiferromagnetic (AFM) order parameter and uniform $d$-wave pairing order. Our stripe wavefunction includes inhomogeneous state with both spin and charge density modulations. } DQMC is a numerically exact auxiliary field method that solves finite-size clusters within the grand canonical ensemble. We use the technique as outlined in Ref.~\cite{JohnstonPRB2013}, applied to rectangular $N = 16\times 4$ clusters with PBC. Additional details of our specific simulation parameters for both methods are provided in the Supplementary Materials (SM) \cite{SOM}. \\

\noindent\textbf{Stripe order at zero temperature} ---  We first examine the static stripe correlations in the Hubbard-Holstein model at $T = 0$ using VMC, focusing on the $\langle n \rangle = 0.875$, $t^\prime = -0.25t$, and $U = 8t$ case. (Additional results for $t^\prime = 0$ are also shown here and in the SM~\cite{SOM}.) For these parameters, we find that a long-range static stripe order produces a lower estimate for the ground state energy than a uniform state. For example, Figs.~\ref{fig:Stripes_VMC}a and \ref{fig:Stripes_VMC}b plot the expectation values of the local staggered spin $S_i^\mathrm{stag} = (-1)^{i_x+i_y}\langle \hat{S}_i^z\rangle$ and local density $\rho_i = \langle \sum_\sigma n_{i,\sigma}\rangle$ operators, respectively, obtained from our optimized variational state when the $e$-ph coupling $\lambda = 0$. The results reveal the typical intertwined unidirectional spin and charge stripe observed in the cuprates, where antiphased regions of antiferromagnetic (AFM) ordered spins are separated by vertical hole-rich regions. For this value of $t^\prime$, the spin and charge modulations have periods $\frac{1}{2}\lambda_\mathrm{spin} = \lambda_\mathrm{charge} \approx 4a$. We also find that the stripe solution is lower in energy for $t^\prime = 0$, but with a different period $\frac{1}{2}\lambda_\mathrm{spin} = \lambda_\mathrm{charge} \approx 8a$, see SM~\cite{SOM}. These results are in agreement with a prior VMC study~\cite{IdoPRB2018}. 

In the limit of $\Omega \rightarrow \infty$, the Holstein model can be mapped onto an effective, attractive Hubbard model with $U = -\lambda W$, reflecting the effective, attractive $e$-$e$ interaction mediated by the lattice. Therefore, for large $\Omega$, we can crudely estimate the effects of the $e$-ph coupling by replacing $U \rightarrow U_\mathrm{eff} = U-\lambda W$. For smaller $\Omega$, this picture still provides a valuable guide for qualitatively understanding the physics of the model; however, additional retardation effects can play a role~\cite{NowadnickPRL2012, JohnstonPRB2013, PhysRevB.98.085405}. Based on these considerations, we naively expect $\lambda \ne 0$ will suppress any correlation-driven phenomena for large $\Omega$. Indeed, prior theoretical work at half-filling has shown that the line $\lambda W \approx U$ defines an approximate boundary between AFM and CDW phases \cite{BauerPRB2010, NowadnickPRL2012, JohnstonPRB2013, KarakuzuPRB2017, Costa2020, PhysRevB.98.085405}.

With this picture in mind, we now examine the influence of the $e$-ph coupling on the static stripes. Figs.~\ref{fig:Stripes_VMC}(c) and \ref{fig:Stripes_VMC}(e) plot $S_i^\mathrm{stag}$ and $\rho_i$ along the ${\bf R}_i = a(i_x, 0)$ direction for representative values of $\lambda = 0,~0.5$ and $\Omega = 5t$. (The data for the other values of $\Omega$ are similar and provided in the SM.) A non-zero coupling $\lambda \ne 0$ reduces the magnitude of both modulations as is evident in the raw VMC data, plotted here as the data points. To quantify this observation across our entire set of simulations, we fit the VMC data with sinusoidal functions, as exemplified by the solid lines. The fitted amplitudes for $S^\mathrm{stag}_i$ and $\rho_i$ are plotted in Figs.~\ref{fig:Stripes_VMC}(d) and \ref{fig:Stripes_VMC}(f), respectively, where results for $t^\prime = -0.25t$ are shown in the main panel and results for $t^\prime = 0$ are shown in the insets.

For both values of $t^\prime$, we find that the Holstein interaction does not change the underlying period of the stripes, but it does affect the amplitudes. For example, increasing $\lambda$ for $t^\prime = -0.25t$ suppresses both the spin and charge modulations and the rate of suppression increases as the phonon frequency $\Omega$ increases. We find qualitatively similar behavior for large $\Omega$ when $t^\prime = 0$, as shown in the insets. However, for lower energy phonons ($\Omega \lesssim 2t$), the spin modulations are weakly suppressed when $\lambda \le 0.75$ while the amplitude of the density modulations \emph{increases}. These results indicate that a Holstein coupling to high-energy phonons suppress static stripes while coupling to lower energy optical modes can stabilize them by enhancing their charge modulations, depending on the value of $t^\prime$. 

\begin{figure}[t]
    \centering
    \includegraphics[width=\columnwidth]{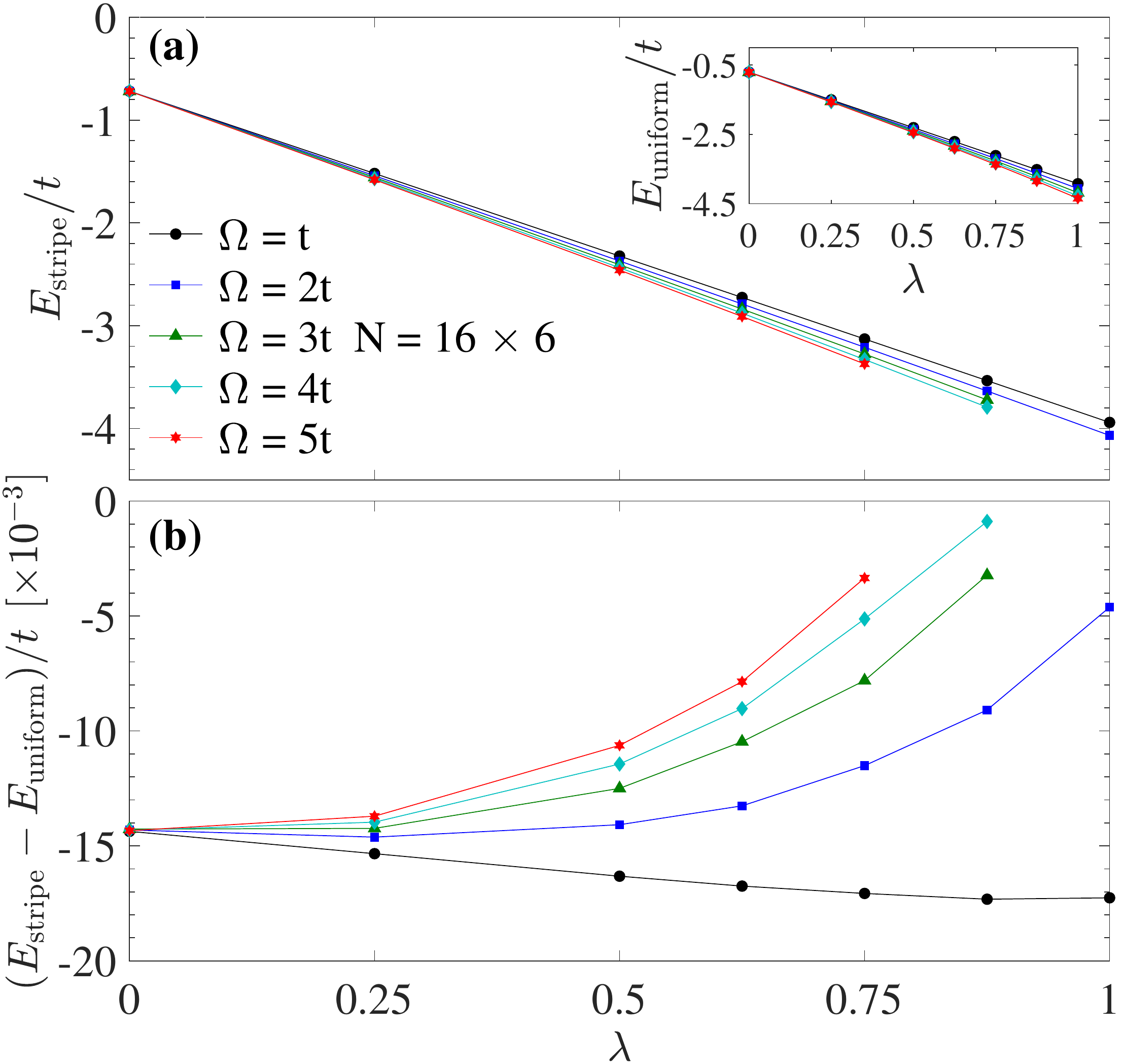}
    \caption{Varational Monte Carlo estimates for the energies of the two-dimensional 
    Hubbard-Holstein model with $t^\prime = 0$ and various phonon energies $\Omega$ as a function of $\lambda$. Panel (a) shows the expectation value of the Hamiltonian $E_\mathrm{stripe}$ for a state with static spin and stripe correlations. The inset shows the expectation value of the Hamiltonian $E_\mathrm{uniform}$ for a state with uniform $d$-wave pairing and AFM correlations. Panel (b) plots the difference 
    $E_\mathrm{stripe}-E_\mathrm{uniform}$ as a function of $\lambda$. All energies are 
    reported in units of $t$.}
    \label{fig:VMC_Summary}
\end{figure}

\begin{figure}[t]
    \centering
    \includegraphics[width=\columnwidth]{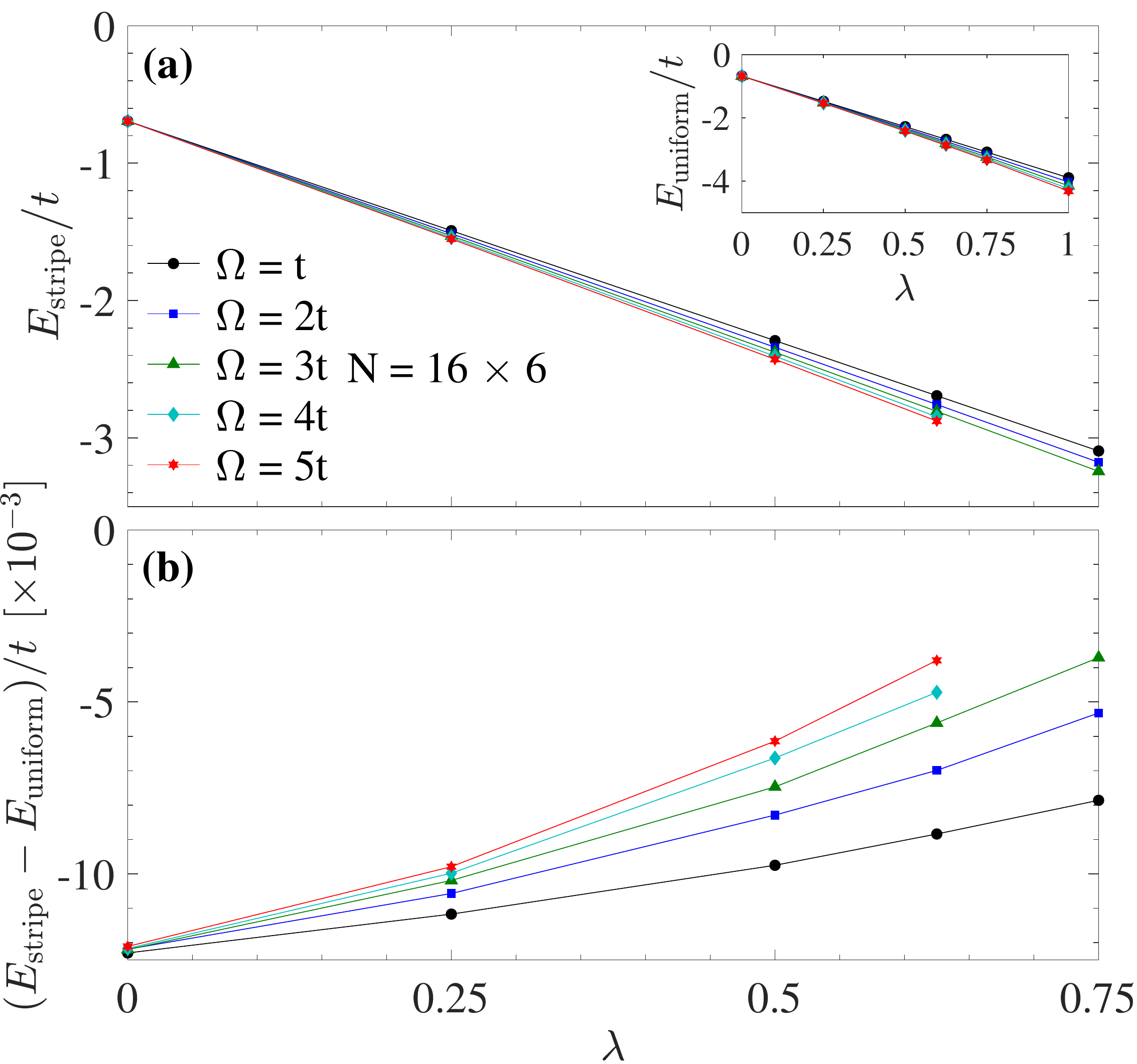}
    \caption{Varational Monte Carlo estimates for the energies of the two-dimensional 
    Hubbard-Holstein model with $t^\prime = -0.25t$ and various phonon energies $\Omega$ as a function of $\lambda$. Panel (a) shows the expectation value of the Hamiltonian $E_\mathrm{stripe}$ for a state with static spin and stripe correlations. The inset shows the expectation value of the Hamiltonian $E_\mathrm{uniform}$ for a state with uniform $d$-wave pairing correlations and AFM order. Panel (b) plots the difference 
    $E_\mathrm{stripe}-E_\mathrm{uniform}$ as a function of $\lambda$. All energies are 
    reported in units of $t$.}
    \label{fig:VMC_Summary2}
\end{figure}

Having assessed the effects of the $e$-ph coupling on stripe correlations, we now turn to the relative energies of the stripe and superconducting states. Figs.~\ref{fig:VMC_Summary} and \ref{fig:VMC_Summary2} compare the variational energies of the stripe and uniform solutions for $t^\prime = 0$ and  $-0.25t$, respectively. Focusing first on the $t^\prime = 0$  case, 
Fig.~\ref{fig:VMC_Summary}(a) plots the expectation value of the system's total energy as a function of $\lambda$ for the optimized variational stripe states, like those shown in Fig.~\ref{fig:Stripes_VMC}. The stripe state's energy decreases approximately linearly with $\lambda$ for all $\Omega$, but the rate of decrease is higher for larger phonon energies. The inset of Fig.~\ref{fig:VMC_Summary}(a) shows the corresponding data for a uniform state, where we observe similar behavior. 

The dominant contribution to change in energy can be attributed to the shift in the lattice's equilibrium position when $\lambda \ne 0$ \cite{JohnstonPRB2013}, which lowers the energies of the uniform and stripe states by a comparable amount. More subtle differences are observed when we examine the energy differences between the two states, as shown in Fig.~\ref{fig:VMC_Summary}(b). When $\lambda = 0$, we find that the stripe state is $0.01437t$/site lower in energy than the uniform state, such that the stripe solution is a better approximation for the ground state at this doping, consistent with Refs. \cite{PhysRevB.97.045138} and \cite{sorella2021phase}. For $\Omega > 2t$, increasing $\lambda$ reduces the energy difference between these states, signaling that the stripe state is suppressed relative to the uniform superconducting state. Conversely, for $\Omega = t$, increasing $\lambda$ stabilizes the stripe order. Interestingly, for intermediate $\Omega = 2t$, we observe nonmonotonic behavior, where the stripe correlations are enhanced at small $\lambda < 0.5$ but suppressed for larger values.

The results for $t^\prime = -0.25t$, shown in Fig.-\ref{fig:VMC_Summary2}, show that the $e$-ph coupling lowers the energy difference between the stripe and uniform solutions at all $\Omega$. However, the effect is reduces as the phonon energy decreases. 

The results in Figs.~\ref{fig:Stripes_VMC}-\ref{fig:VMC_Summary2} suggest that anti-adiabatic phonons destabilize static stripes relative to the uniform state by suppressing both the spin and charge modulations. However, lower-energy phonons can have the opposite effect, increasing the strength of charge modulations while leaving the spin modulations relatively unchanged when  $t^\prime = 0$. While we can understand the large $\Omega$ behavior in terms of a reduced $U_\mathrm{eff}$, the small $\Omega$ behavior, which is more relevant for the high-$T_\mathrm{c}$ cuprates, highlights the critical role of retardation. For example, studies examining the competition between AFM and CDW in the half-filled Hubbard-Holstein model generally find that the AFM correlations can persist beyond the line defined by $U = \lambda W$ when $\Omega \le t$. In this regime, the lattice cannot respond quickly enough to the double occupations created in the AFM exchange process and is less effective in suppressing the AFM correlations. A similar effect appears to be at play here, as the AFM spin stripe correlations are less affected by the $e$-ph interaction as $\Omega$ decreases. At the same time, smaller values of $\Omega$ correspond to softer springs, which means that the hole-rich regions can lower their energy more readily by forming large local lattice distortions for a fixed value of $\lambda$. These combined effects provide a plausible explanation for our VMC results. \\

\begin{figure}[t]
    \centering
    \includegraphics[width=0.8\columnwidth]{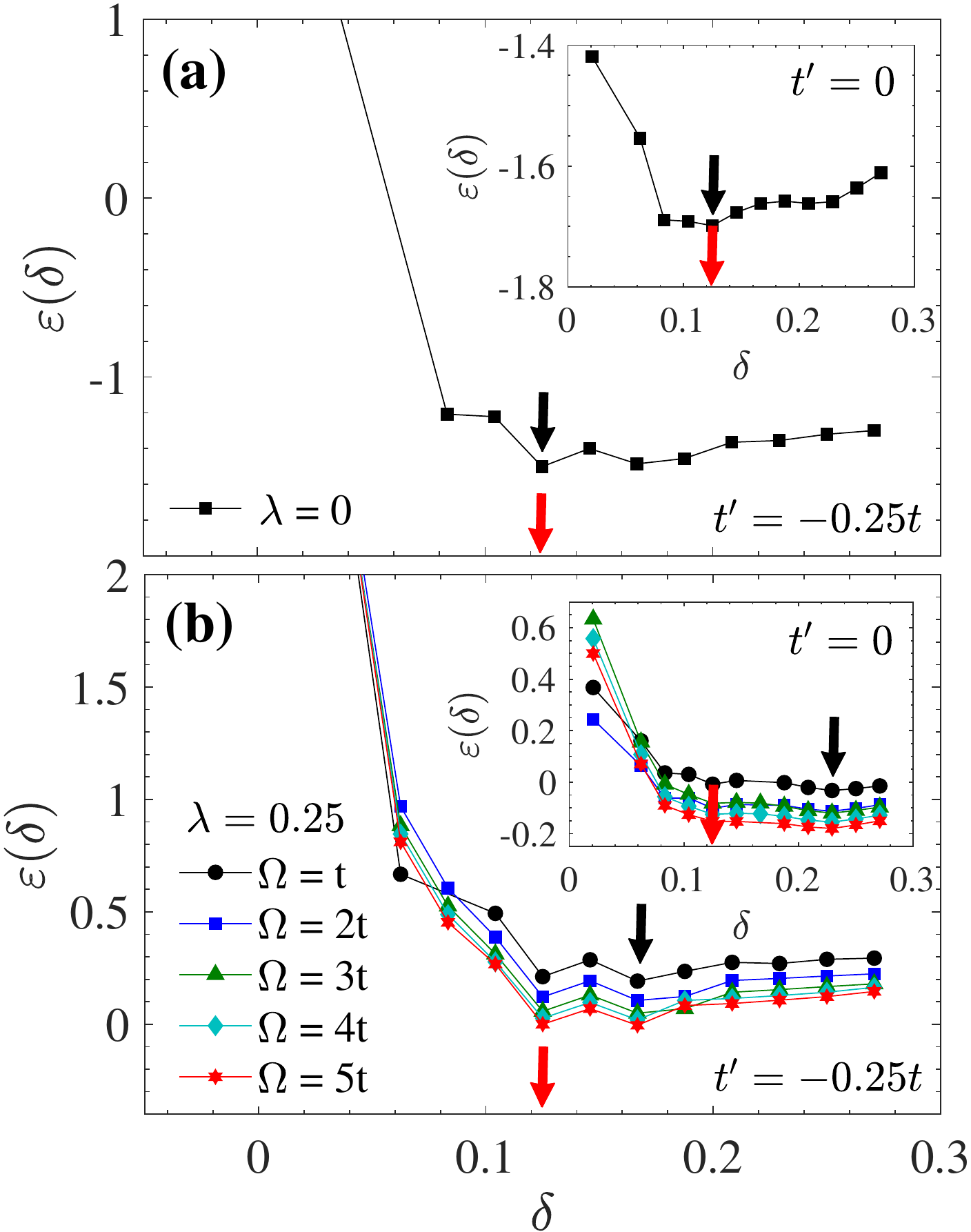}
    \caption{The energy per hole $\varepsilon(\delta)$ for $t^\prime= -0.25 t$ and $t^\prime= 0$ (insets) at different frequencies. The red arrows indicate the doping level at which VMC was performed and black arrows indicate the global minimum of $\varepsilon(\delta)$. Panels (a) and (b) show the energy per hole for $\lambda = 0$ and $\lambda = 0.25$, respectively.}
    \label{fig:Phase_Sep}
\end{figure}

\noindent\textbf{Phase separation} --- 
If the system is in a stable phase, then the ground state energy as a function of filling is convex, i.e. $\partial^2E(n)/\partial n^2 > 0$. One can determine when this condition is violated by calculating the energy per hole \cite{PhysRevLett.64.475}
\begin{equation}\label{eq:Eperhole}
\varepsilon(\delta) = \frac{E(\delta)-E(0)}{\delta}, 
\end{equation}
where $\delta = 1 - \langle\hat{n}\rangle$ is the excess hole density. If 
$\epsilon(\delta)$ has a local minimum at some hole doping $\delta_c$, then the system will 
phase separate for any $\delta < \delta_c$. Using this approach, several VMC studies have observed a tendency towards phase separation in the lightly hole doped Hubbard model~\cite{PhysRevLett.64.475,  IdoPRB2018, PhysRevB.94.195126}. For example, Ido \emph{et al}. [Ref.~\cite{IdoPRB2018}] concluded that phase separation occurs for doping levels $\delta < 0.125$ in the singleband model when $U = 10t$ and $t^\prime = 0$. 
But VMC tends to over predict the value of $\delta_c$; this occurs because the division by $\delta$ in Eq.~\eqref{eq:Eperhole} magnifies small errors in the variational energies $E(\delta)$ at low doping levels. Indeed, Ref.~\cite{PhysRevB.94.195126} systematic improvements of the variational energies via Greens function Monte Carlo with the Fixed Node approximation drove critical doping level $\delta_c$ very close to zero. Notably, the nature of the correlations in the ground state did not change as the wave function was improved. 

\begin{figure*}[ht]
    \centering
    \includegraphics[width=0.8\textwidth]{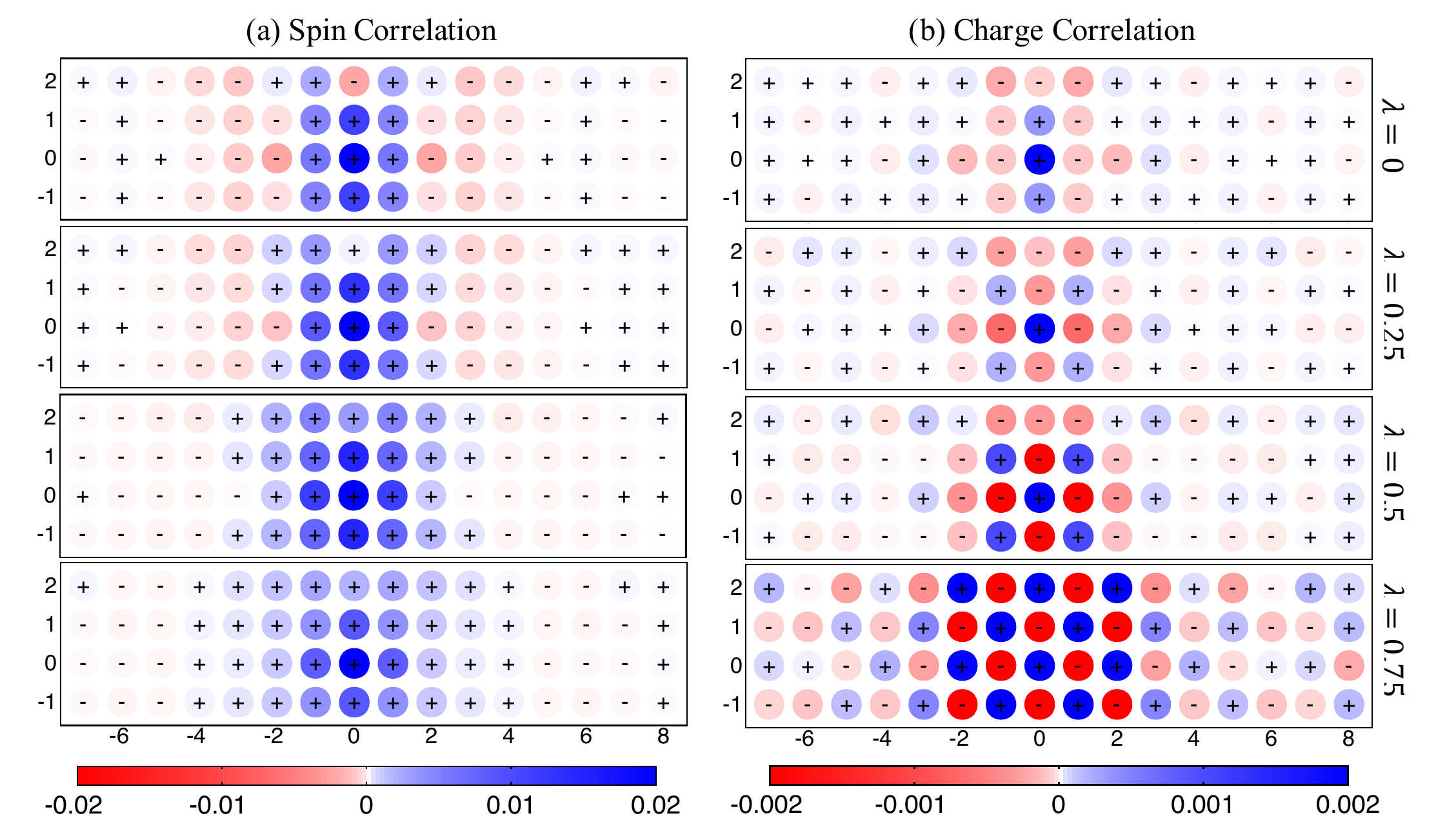}
    \caption{DQMC results for the fluctuating (a) spin and (b) charge stripes in the Hubbard-Holstein model at a filling of $\langle \hat{n} \rangle = 0.8$ and an inverse temperature of $\beta = 4/t$, obtained on $N = 16\times 4$ clusters with $t^\prime= -0.25 t$, $U = 6t$, and $\Omega = t/2$. Each row in (a) shows the real-space static staggered spin-spin correlation function at different values of $\lambda$. The panels in (b) show the corresponding density-density correlation functions.
}
\label{fig:DQMC_rspace}
\end{figure*} 

With these caveats in mind, Fig.~\ref{fig:Phase_Sep} examines $\varepsilon(\delta)$ for our model parameters for completeness. For $\lambda = 0$ [Fig.~\ref{fig:Phase_Sep}(a)], we find that $\varepsilon(\delta)$ 
reaches a local minima at $\delta_c = 0.125$ for both $t^\prime = 0$ and $-0.25t$, in agreement with Ref.~\cite{IdoPRB2018}. Once the $e$-ph interaction is included, the local minima develop at $\delta_c = 0.125$ and $0.167$ for $t^\prime = -0.25t$ and all values of $\Omega$, with the latter being the global minimum. The predicted tendency towards phase separation is more severe for the case of $t^\prime = 0$ with the region extending to $\delta_c = 0.229$, as shown in the inset of Fig~\ref{fig:Phase_Sep}(b).

These results suggest that the Holstein coupling increases the tendency toward phase separation commonly found in the Hubbard and $t$-$J$ models \cite{PhysRevLett.64.475, IdoPRB2018, PhysRevB.94.195126}. The doping level of our simulations, indicated by the red arrows, fall below $\delta_c$ when $\lambda \ne 0$. However, the local minima are largely determined by the rapid growth in $\varepsilon(\delta)$ at low doping, which can be dramatically impacted by small errors in $E(\delta)$. Indeed, Ref.~\cite{PhysRevB.94.195126} found that this rapid rise all but disappears in the $\lambda = 0$ case as the variational  wave functions are improved. We expect a similar change to occur here for $\lambda \ne 0$; however, more detailed simulations will be needed to verify this conjecture. Nevertheless, we believe that we are overestimating $\delta_c$ here, and the actual value is smaller. This conjecture is also supported by the fact that we find no indications for phase separation in our finite temperature DQMC results discussed in the next section.\\

\noindent\textbf{Stripe correlations at finite temperature} --- We now turn to the influence of the Holstein coupling on the fluctuating stripe correlations observed by DQMC at finite temperatures. Fluctuating stripes are challenging to resolve in the Hubbard model at high-temperatures. For this reason, we focused on $N = 16\times 4$ clusters with $\langle \hat{n} \rangle = 0.8$ with $t^\prime = -0.25t$, $U = 6t$, and $\beta = 4/t$, where the fluctuating spin stripe correlations are particularly strong~\cite{HuangQuantMat2018}. Fig.~\ref{fig:DQMC_rspace}(a) plots real-space staggered static spin-spin correlation function 
\begin{equation*}
    C_\mathrm{spin}^\mathrm{stag}({\bf r}_i) =\frac{1}{N} \int_0^\beta  (-1)^{i_x+i_y}\langle \sum_j \hat{S}_{i+j}^z (\tau)\hat{S}^z_{j}(0)\rangle d\tau,
\end{equation*}
measured in DQMC simulations for several values of $\lambda$ and $\Omega = t/2$. (Results for $\Omega = t$ and $\Omega = 2t$ are provided in the SM~\cite{SOM}.) Fig.~\ref{fig:DQMC_rspace}(b) shows the corresponding static density-density correlation function 
\begin{equation*}
    C_\mathrm{den}({\bf r}_i) =\frac{1}{N} \int_0^\beta \sum_j \left[ \langle \hat{n}_{i+j}(\tau) \hat{n}_{j}(0)\rangle - \langle \hat{n}_{i+j}(\tau)\rangle\langle \hat{n}_{j}(0)\rangle \right]d\tau. 
\end{equation*}

For $\lambda = 0$ [top row, Figs.~\ref{fig:DQMC_rspace}(a),(b)], we observe fluctuating spin and charge stripe correlations consistent with prior work~\cite{HuangQuantMat2018,HuangPreprint,mai2021stripes}. The spin correlations are evident from the short-range AFM correlations (blue) surrounded by other AFM domains where the magnetic correlations are flipped (red). At this temperature, the charge stripes are weak but still present. 

As with the VMC results for this value of $t^\prime$, we find that the spin correlations are reduced when we introduce the $e$-ph coupling. This suppression occurs gradually for weak coupling, but it accelerates as $\lambda$ increases. For example, for $\lambda = 0.5$, we already see that the red domains in $C_\mathrm{spin}^\mathrm{stag}({\bf r}_i)$ are nearly absent while the AFM correlations persist over longer length scales. As $\lambda$ increases further, the AFM correlations extend over a larger distance. This behavior is presumably due to the attractive interaction mediated by the phonons, which counteracts the correlations generated by the Hubbard $U$. In contrast to the spin correlations, the charge correlations remain more robust for $\lambda \le 0.5$ but begin to develop an additional short-range ${\bf Q} = (\pi,\pi)$ modulation. These new modulations signal the formation of local bipolarons~\cite{nosarzewski2021superconductivity}, which tend to arrange themselves in local checkerboard-like order. Finally, the $(\pi,\pi)$ modulation dominates over the entire cluster for the largest value of $\lambda$ studied here.

\begin{figure*}[ht]
    \centering
    \includegraphics[width=\textwidth]{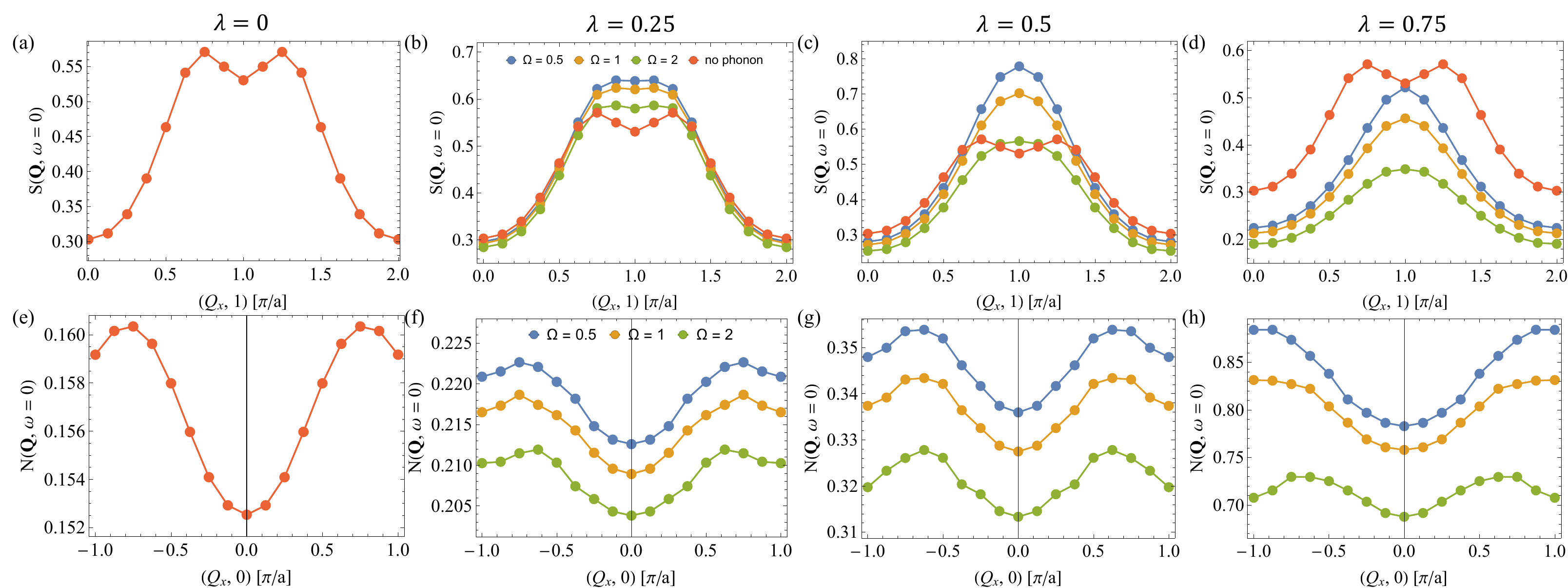}
    \caption{DQMC results for the fluctuating spin and charge stripes of the Hubbard-Holstein model in momentum space. All results were obtained at  a filling of $\langle \hat{n} \rangle = 0.8$ and an inverse temperature of $\beta = 4/t$ using $N = 16\times 4$ clusters with $t^\prime= -0.25 t$ and $U = 6t$. The top row shows the static spin susceptibility for different values of $\lambda$. The bottom row shows the corresponding charge susceptibility. Results are shown for $\Omega = t/2$, $t$, and $2t$, as indicated in the legends provided in the second column.
}
\label{fig:DQMC_kspace}
\end{figure*}

So far, we have focused on results for $\Omega = t/2$; however, we have obtained similar results for $\Omega = t$ and $2t$ (see also \cite{SOM}). The trends across the data set are more easily summarized by examining the momentum-dependent static spin $S({\bf Q}, \mathrm{i}\omega_n=0)$ and charge $N({\bf Q}, \mathrm{i}\omega_n=0)$ susceptibilities, shown in Fig.~\ref{fig:DQMC_kspace}. These quantities are obtained by Fourier transforming the unequal time spin-spin and density-density correlation functions and integrating over imaginary time.

The spin stripe correlations manifest in $S({\bf Q},\mathrm{i}\omega_n=0)$ as incommensurate peaks at ${\bf Q} = (\pi \pm \delta_s,\pi)$. As with the real-space picture, we find that $\delta_s$ is reduced for small $\lambda$, causing the double peaks to merge into a single broad peak centered at $(\pi,\pi)$. However, in most cases, we can still discern two components by fitting a set of lorentzian functions to the data (see the SM~\cite{SOM}). For large $\lambda$, $S({\bf Q},0)$ approaches a single AFM peak, which is suppressed as $\Omega$ increases due to the reduction of $U_\mathrm{eff}$ and increasing competition with the competing CDW phase. 

The charge stripes manifest in $N({\bf Q},i\omega_n=0)$ as incommensurate peaks centered at ${\bf Q} = (\delta_c,0)$~\cite{mai2021stripes}. In the absence of $e$-ph coupling, we clearly observe this structure with $\delta_s/\delta_c \approx 0.45$ at $\beta = 4/t$, consistent with Ref.~\cite{mai2021stripes}. These peaks remain well defined for $\lambda \le 0.5$ for all values of $\Omega$, while the overall magnitude of the charge correlations increases uniformly with decreasing $\Omega$. It is only for stronger $e$-ph coupling ($\lambda = 0.75$) that $\delta_c$ is shifted towards $\pi$ as the phonon frequency is decreased, signalling a suppression of the charge stripes. We also note that increasing $\lambda$ enhances the ${\bf  Q} = (\pi,\pi)$ charge correlations~\cite{SOM} due an increased tendency towards bipolaron formation. In fact, the $(\pi,\pi)$ correlations become significantly larger than the $(\delta_c,0)$ ones if $\lambda$ is too large, consistent with the real space picture shown in Fig~\ref{fig:DQMC_rspace}. 

The overall picture obtained from our DQMC results ($t^\prime =-0.25t$) is that the Holstein interaction tends to suppress the fluctuating stripe correlations but that the effect is more pronounced for the spin correlations at small $\Omega$ and weak to intermediate coupling. For strong coupling and/or large $\Omega$, the $e$-ph interaction suppresses both channels due to the competition with $(\pi,\pi)$ charge correlations, bipolaron tendencies, or an overall reduction in $U_\mathrm{eff}$. These results are in agreement with our zero-temperature VMC results for $t^\prime = -0.25t$. These results demonstrate that the $e$-ph interaction can have a non-trivial effect on static and fluctuating stripe correlations.\\

\noindent\textbf{Discussion} --- Several numerical methods have found evidence that a Holstein interaction can enhance $d$-wave pairing correlations in the doped singleband Hubbard-Holstein model \cite{HuangPRB2003, PhysRevB.75.014503, MendlPRB2017}. Some researchers have linked this enhancement to a non-trivial 
screening of the $e$-ph interaction, which reduces large-${\bf q}$ scattering relative to small ${\bf q}$ \cite{HuangPRB2003, JohnstonPRL2012, JohnstonPRB2010}. Our results suggest that the lattice can affect superconductivity in another way by altering the competing stripe phases.   
Different optical phonon branches can play different roles in this context: high-energy phonons generally suppress both spin and charge stripes. In contrast, low-energy phonons can enhance them by increasing the charge modulations. Since cuprate optical oxygen phonons have $\Omega \leqslant t/3$, we expect that the latter regime is more relevant to these materials. Our results, therefore, provide a natural framework for understanding anomalous oxygen isotope effects observed in materials like La$_{2-x}$Sr$_x$CuO$_4$ \cite{Isotope}. They also suggest an explanation for why charge modulations tend to develop before spin modulations in the cuprates, even though fluctuating spin stripes appear to be stronger in the doped Hubbard model. 

Why does the $e$-ph interaction primarily couple to the charge modulations? To some extent, this might be expected since the phonons couple directly to the local charge density. There are other factors to consider, however. For one, stripe formation is a form of phase separation that emerges as a compromise in balancing the doped carriers' kinetic and potential energies. Several QMC studies of the Holstein model have found evidence that it is prone to phase separation when doped away from half-filling \cite{KarakuzuPRB2017, paleari2021quantum, bradley2020superconductivity}, which is stronger for low-energy phonons at high temperatures \cite{KarakuzuPRB2017, bradley2020superconductivity}. An enhancement of the $e$-ph coupling at small-{\bf q} can further exacerbate this tendency \cite{PhysRevB.99.205145, PhysRevB.99.075108}. In the case of the doped Hubbard-Holstein model, these factors then cooperate in collecting the doped carriers into particular spatial regions of the system, which would explain our observations. 

Many models and experimental measurements on the cuprates estimate the dimensionless $e$-ph coupling to the oxygen-derived phonon modes to be in the range $\lambda \sim 0.3-1$ \cite{LanzaraNature2001, PhysRevLett.93.117004, ShenPRL2004, LeeNature2006, JohnstonPRL2012, RossiPRL2019}. Combined with our results, these estimates suggest that the $e$-ph coupling could play a role in shaping the stripe correlations. The relevant phonon frequencies in the real materials ($\Omega \approx t/3$) are smaller than the values considered here. With this in mind, our results for $t^\prime = -0.25t$ suggest that a Holstein interaction will be insufficient for stabilizing the superconducting state. However, more sophisticated $e$-ph models must also be examined before we can draw definitive conclusions about the real materials. For example, coupling to the bond-buckling modes occurs via the oxygen on-site (potential) energy and has a significant momentum dependence \cite{PhysRevLett.93.117004, JohnstonPRB2010}, unlike the Holstein model studied here. Similarly, coupling to the bond-stretching modes occurs via a Su-Schrieffer-Heeger (SSH) type interaction \cite{PhysRevLett.93.117004, JohnstonPRB2010}, which modulates carrier's kinetic energy~\cite{LiQM2020, PhysRevLett.121.247001}. To fully understand the role of these interactions on cuprate stripes, it will be necessary to study generalizations of the three-band model, which can capture these aspects of the relevant phonon modes. Nevertheless, our results demonstrate that $e$-ph coupling cannot be neglected in any complete picture of stripe physics. 

\vspace{0.5cm}
\begin{acknowledgments}
\noindent\emph{Acknowledgements} --- This work was supported by the U.S. Department of Energy, Office of Science, Office of Basic Energy Sciences, under Award Number DE-SC0022311. The DQMC calculations used the Extreme Science and Engineering Discovery Environment (XSEDE) expanse supercomputer \cite{XSEDE} through the startup allocation TG-PHY210057, which is supported by National Science Foundation grant number ACI-1548562. 
\end{acknowledgments}

\bibliography{references}
\end{document}


\title{Supplementary materials for ``Stripe correlations in the two-dimensional Hubbard-Holstein model''}
\date{\today}

\author{Seher Karakuzu}
\affiliation{Computational Sciences and Engineering Division, Oak Ridge National Laboratory, Oak Ridge, TN 37831-6164, USA\looseness=-1}
\affiliation{Center for Computational Quantum Physics, Flatiron Institute, 162 5th Avenue, New York, NY 10010, USA\looseness=-1}

\author{Andy Tanjaroon Ly}
\affiliation{Department of Physics and Astronomy, The University of Tennessee, Knoxville, TN 37966, USA}

\author{Peizhi Mai}
\affiliation{Computational Sciences and Engineering Division, Oak Ridge National Laboratory, Oak Ridge, TN 37831-6164, USA\looseness=-1}
\affiliation{Department of Physics and Institute of Condensed Matter Theory, University of Illinois at Urbana-Champaign, Urbana, IL 61801, USA}

\author{James Neuhaus}
\affiliation{Department of Physics and Astronomy, The University of Tennessee, Knoxville, TN 37966, USA}

\author{Thomas A. Maier}
\affiliation{Computational Sciences and Engineering Division, Oak Ridge National Laboratory, Oak Ridge, TN 37831-6164, USA\looseness=-1}

\author{Steven Johnston}
\affiliation{Department of Physics and Astronomy, The University of Tennessee, Knoxville, TN 37966, USA}
\affiliation{Institute for Advanced Materials and Manufacturing, University of Tennessee, Knoxville, TN 37996, USA\looseness=-1}

{
\let\clearpage\relax
\maketitle
}
\large
\noindent\textbf{Supplementary Note 1: Details of the VMC calculations}\normalsize

The trial wave function we used in the VMC calculations are parameterized as 
\begin{equation}\label{eq:wf}
|\Psi_T\rangle = {\cal J}_d^{ee} {\cal J}_s^{ee} {\cal J}^{ep} {\cal P}({N_e}) |\Phi_T^{el}\rangle \otimes |\Phi_T^{ph}\rangle.
\end{equation}
Here, $|\Phi_T^{el}\rangle$ is the ground state of a quadratic mean-field Hamiltonian 
\begin{equation}
 {\cal H}_{\rm MF} = {\cal H}_{\rm 0} + {\cal H}_{\rm AFM} + {\cal H}_{\rm SC} + {\cal H}_{\rm stripe}^c + {\cal H}_{\rm stripe}^s, 
\label{eq:auxham}
\end{equation}
which contains electron hopping, antiferromagnetism (AFM), $d$-wave singlet pairing, and spin an charge stripes as follows
\begin{equation}
 {\cal H}_{\rm 0} = -t \sum_{\langle i,j \rangle,\sigma} c^\dag_{i,\sigma} c^{\phantom\dagger}_{j,\sigma} + {\rm h.c.}
- \mu \sum_{i,\sigma} c^\dag_{i,\sigma} c^{\phantom\dagger}_{i,\sigma}, 
\label{eq:h0}
\end{equation}

\begin{equation}
 {\cal H}_\mathrm{AFM} = \Delta_{\rm AFM}  \sum_{i} (-1)^{i_x + i_y} (c^\dag_{i,\uparrow}c^{\phantom\dagger}_{i,\uparrow} - c^\dag_{i,\downarrow}c^{\phantom\dagger}_{i,\downarrow}  ), 
\label{eq:haf}
\end{equation}

\begin{equation}
 {\cal H}_\mathrm{d-wave} =\sum_{\langle i,j \rangle} \Delta_{\rm i,j}  ( c^\dag_{i,\uparrow} c^\dag_{j,\downarrow} + c^\dag_{j,\uparrow} c^\dag_{i,\downarrow} + {\rm h.c.} ), 
\label{eq:Hsc}
\end{equation}


\begin{equation}
 {\cal H}_{\rm stripe}^{s} =   \sum_{i} \Delta^{\rm s}_{i_x} (-1)^{i_x + i_y} (c^\dag_{i,\uparrow}c^{\phantom\dagger}_{i,\uparrow} - c^\dag_{i,\downarrow}c^{\phantom\dagger}_{i,\downarrow}  ),~\mathrm{and}
\label{eq:Hdstripe}
\end{equation}

\begin{equation}
 {\cal H}_{\rm stripe}^{c} =   \sum_{i} \Delta^{\rm c}_{i_x}  (c^\dag_{i,\uparrow}c^{\phantom\dagger}_{i,\uparrow} + c^\dag_{i,\downarrow}c^{\phantom\dagger}_{i,\downarrow}  ). 
\label{eq:Hcstripe}
\end{equation}
The superconducting order parameter in Eq.~(\ref{eq:Hsc}) is defined as $\Delta_{i,j} = \Delta_\mathrm{d-wave}[\delta_{i+\hat{x},j} 
+ \delta_{i-\hat{x},j} - \delta_{i+\hat{y},j} 
-\delta_{i-\hat{y},j}]$. 
In Eqs.~(\ref{eq:h0})--(\ref{eq:Hcstripe}), the variational parameters $\mu$, $\Delta_{\rm AFM}$, $\Delta_{\rm d-wave}$, $\Delta^\mathrm{s}_{i_x}$, and $\Delta^\mathrm{c}_{i_x}$ are optimized to minimize the variational energy using the stochastic reconfiguration method~\cite{Sorella2005}. Here, ${\bf r}_i=a(i_x,i_y)$ with $i_x,~i_y\in \mathbb{Z}$ indicates the coordinates of site $i$ in the square lattice, and $\sigma=+1\, (-1)$ for up (down) electrons. Note that $\Delta^\mathrm{s}_{i_x}$ and $\Delta^\mathrm{c}_{i_x}$ constitute $2\times 16$ variational parameters, one for each column of sites along the long direction of the $16\times 6$ cluster. 

The projector ${\cal P}({N_e})$ is used to fix the total number of electrons $N_e$ in the system. The uncorrelated phononic part in the wave function is then given by
 \begin{equation}
     |\Phi_T^{ph} \rangle = e^{\mu_{ph} \sum_{i} n_{i}^{ph}} e^{b_{k=0}^{\dag}} |0 \rangle, 
 \end{equation}
where $\mu_{ph}$ is the fugacity term which controls the total number of phonons, $n_i^{ph}$  is the total number of phonons on $i$\textsuperscript{th} site, and $|0\rangle$ is the phonon vacuum state.

The terms ${\cal J}_d^{ee}$ and  ${\cal J}_s^{ee}$ are density-density and spin-spin Jastrow correlators for the electron-electron correlations and ${\cal J}^{ep}$ is electron-phonon Jastrow correlator. They are written as 
\begin{eqnarray}
&& {\cal J}_d^{ee} = \exp \left ( -\frac{1}{2} \sum_{i,j} v^{ee}_{i,j} n_{i} n_{j} \right ), \label{eq:Jeed} \\
&& {\cal J}_s^{ee} = \exp \left ( -\frac{1}{2} \sum_{i,j} w^{ee}_{i,j} S_{i} S_{j} \right ),~\mathrm{and}\label{eq:Jees} \\
&& {\cal J}^{ep} = \exp \left ( - \sum_{i,j} v^{ep}_{i,j} n_{i} n_{j}^{ph} \right ), \label{eq:Jep} 
\end{eqnarray}
where $v^{ee}_{i,j}$, $w^{ee}_{i,j}$, and $v^{ep}_{i,j}$ are long ranged variational parameters that enter as pseudo-potentials. Each element is taken to be translationally invariant, such that they depend only upon the distance between $i$\textsuperscript{th} and $j$\textsuperscript{th} sites, i.e., $|{\bf R}_{i}-{\bf R}_{j}|$. Moreover, each of the pseudo-potentials are taken to be symmetric, e.g. $v^{ee}_{i,j} = v^{ee}_{j,i}$, and this is valid for all parameters in all Jastrow correlators. We optimize each of these parameters independently via stochastic reconfiguration method to reach the lowest possible energy, which approximates the ground state of the system. 

We performed VMC calculations on $16\times 6$ clusters with periodic boundary conditions. Parameter optimization was performed for $3\times 10^3$ steps (continued for $1\times 10^4$ steps for $d$-wave) and $1\times 10^6$ Monte Carlo steps. We obtained statistical error using Jack-Knife re-sampling. \\

\large\noindent\textbf{Supplementary Note 2: Additional VMC results}\normalsize

\begin{figure}[ht]
    \centering
    \includegraphics[width=0.9\textwidth]{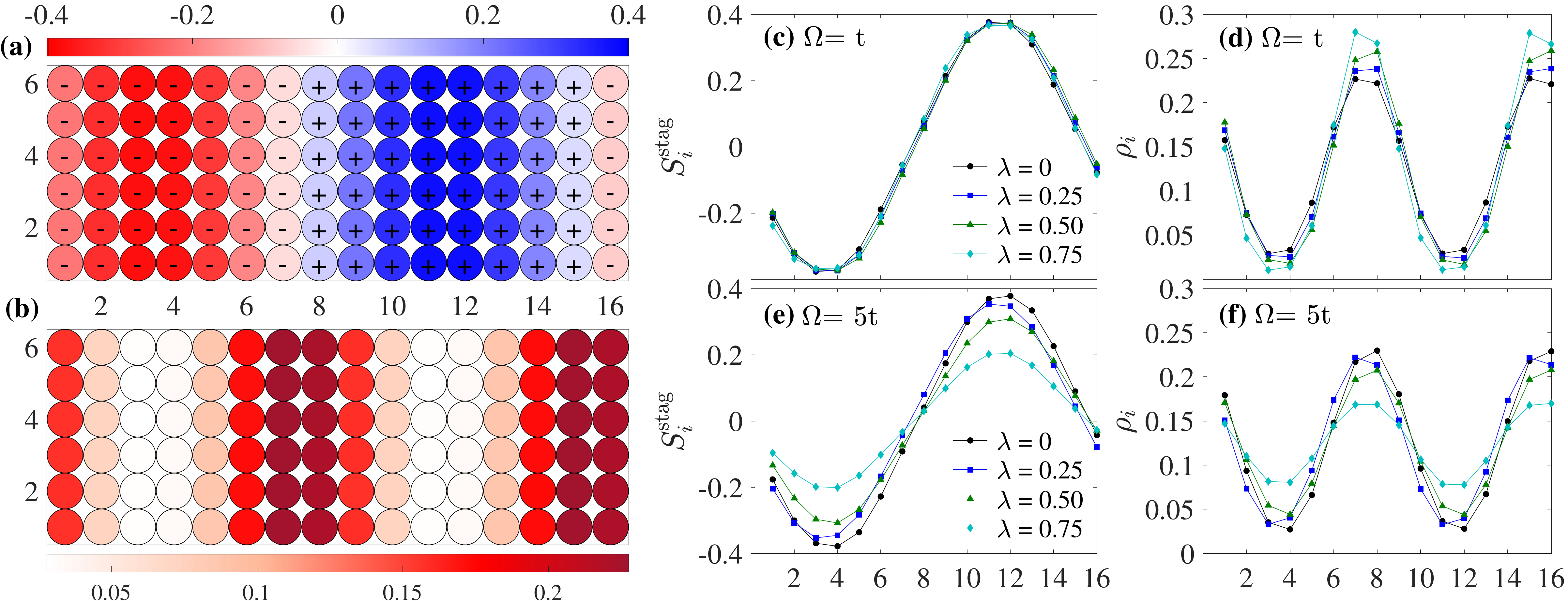}
    \caption{Variational Monte Carlo results for static spin and charge stripes in the Hubbard-Holstein model with $U = 8t$ and $t^\prime = 0$. (a) The expectation value of the staggered local spin operator in real space $S_i^\mathrm{stag} = (-1)^{i_x+i_y}\langle \hat{S}_i^z\rangle$ and $\lambda = 0$, where $S_i^z = \frac{1}{2}(\hat{n}_{i,\uparrow}-\hat{n}_{i,\downarrow})$ is the $z$-component of the local spin operator. (b) The expectation value of the local density operator $\rho_i = \langle \sum_\sigma \hat{n}_{i,\sigma} \rangle$ for the same case. Panels (c) and (d) show $S_i^\mathrm{stag}$ and $\rho_i$, respectively, along the line $(i_x,0)$ for several values of $\lambda$ when $\Omega = t$, while panels (e) and (f) show analogous results for a large $\Omega = 5t$. }
    \label{fig:VMC_Maps_tpeq0}
\end{figure}

\begin{figure}[ht]
    \centering
    \includegraphics[width=0.9\textwidth]{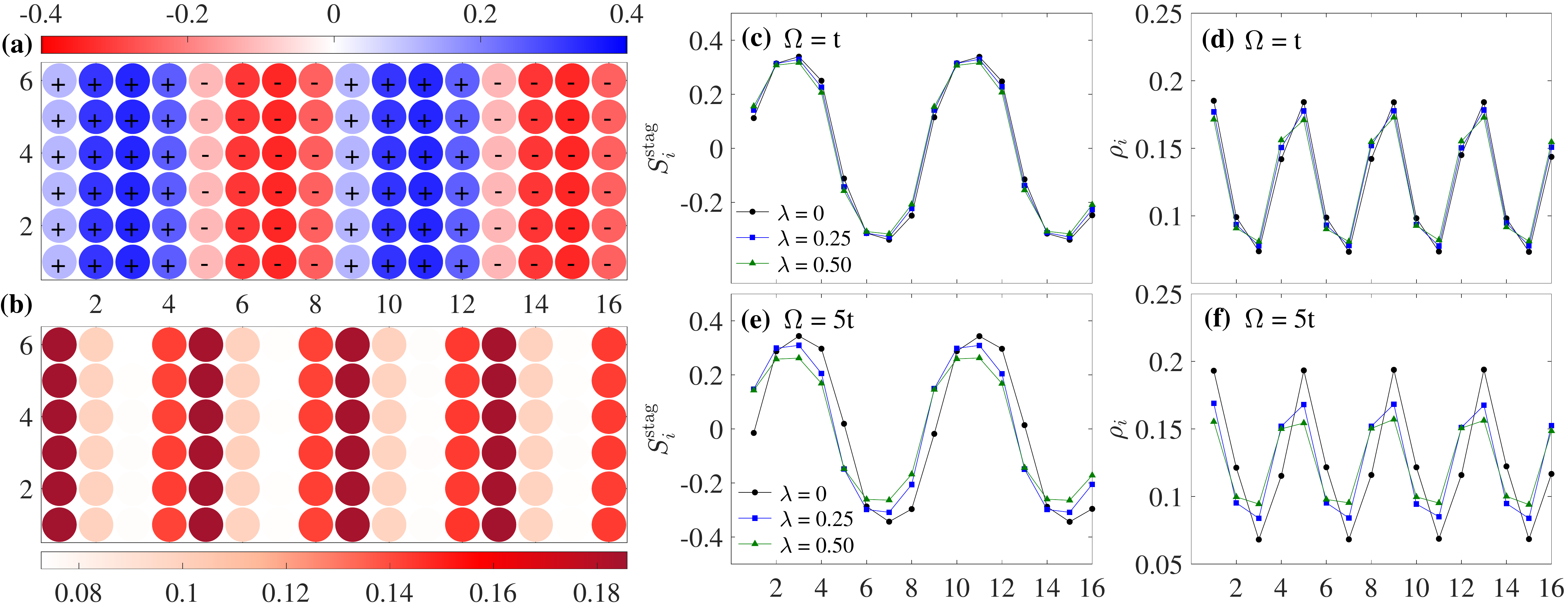}
    \caption{Variational Monte Carlo results for static spin and charge stripes in the Hubbard-Holstein model with $U = 8t$ and $t^\prime = -0.25t$. (a) The expectation value of the staggered local spin operator in real space $S_i^\mathrm{stag} = (-1)^{i_x+i_y}\langle \hat{S}_i^z\rangle$ and $\lambda = 0$, where $S_i^z = \frac{1}{2}(\hat{n}_{i,\uparrow}-\hat{n}_{i,\downarrow})$ is the $z$-component of the local spin operator. (b) The expectation value of the local density operator $\rho_i = \langle \sum_\sigma \hat{n}_{i,\sigma} \rangle$ for the same case. Panels (c) and (d) show $S_i^\mathrm{stag}$ and $\rho_i$, respectively, along the line $(i_x,0)$ for several values of $\lambda$ when $\Omega = t$, while panels (e) and (f) show analogous results for a large $\Omega = 5t$. 
    }
    \label{fig:VMC_Maps_tpeqm0p25}
\end{figure}

\begin{figure}[h]
    \centering
    \begin{minipage}{0.49\textwidth}
    \includegraphics[width=\textwidth]{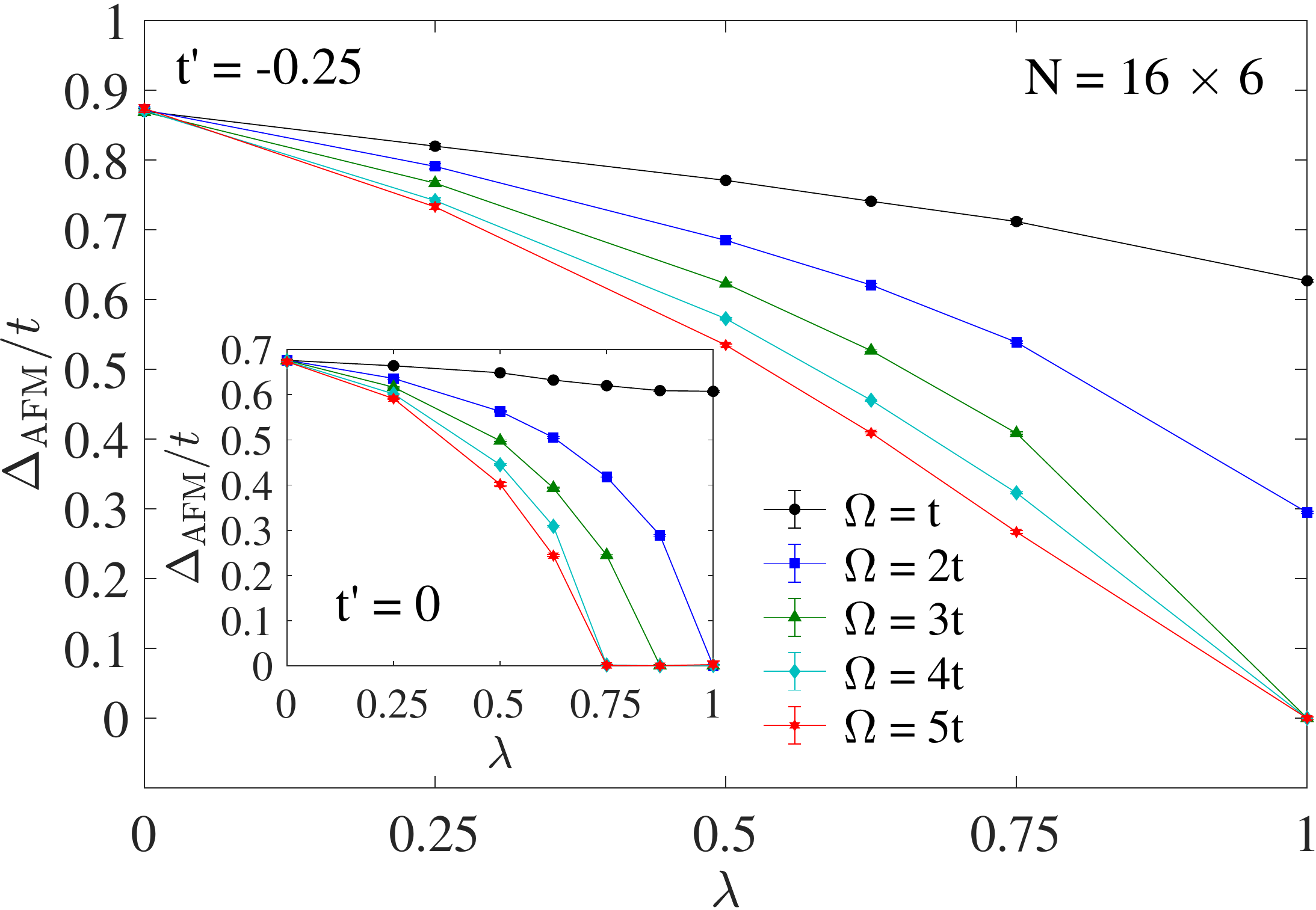}
    \end{minipage}
    \hfill
    \begin{minipage}{0.49\textwidth}    
    \includegraphics[width=\textwidth]{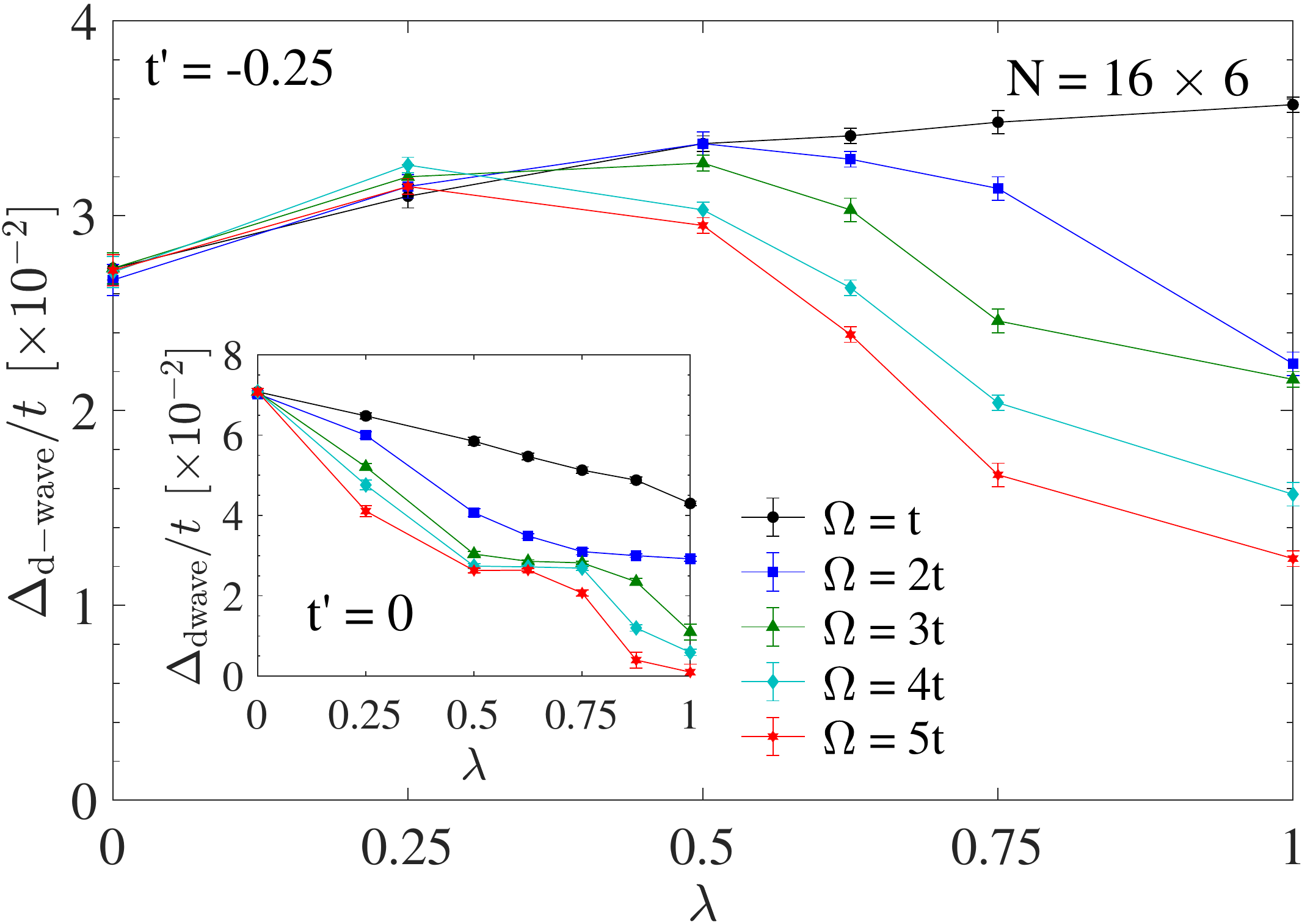}
    \end{minipage}
    \caption{The evolution of the antiferromagnetic (left) and $d$-wave superconducting order parameters used to define the variational 
    wavefunctions shown in the main text as a 
    function of $\lambda$.} \label{fig:VMC_OP}
\end{figure}

Figures~\ref{fig:VMC_Maps_tpeq0} and ~\ref{fig:VMC_Maps_tpeqm0p25} summarize our VMC results for the evolution of the stripe correlations as a function of $\lambda$ and $\Omega$. Results are shown for $t^\prime = 0$ and $t^\prime = -0.25t$ in Figs.~\ref{fig:VMC_Maps_tpeq0} and ~\ref{fig:VMC_Maps_tpeqm0p25}, respectively. The evolution of the antiferromagnetic and $d$-wave superconducting order parameters used to define the uniform variational wavefunctions shown in the main text are plotted in Fig.~\ref{fig:VMC_OP}. 

\clearpage
\noindent\large\textbf{Supplementary Note 3: Details of the DQMC simulations} \normalsize

We performed DQMC simulations on $16\times 4$ clusters with periodic boundary conditions. The value of $\Delta \tau$ was fixed to $0.1/t$. Measurements were averaged over $1080$ independent Markov chains, each performing $5\times 10^4$ warm-up and $4\times 10^5$ measurement sweeps to generate $10$ bins of data/Markov chain. The measurement is made at every 4th sweep. We obtained error estimates using Jack-Knife re-sampling.  

The momentum-dependent spin susceptibility 
$S({\bf Q},\mathrm{i}\omega_n)$ is obtained by measuring and Fourier transforming the unequal-time 
spin-spin correlation function \cite{NowadnickPRL2012} 
\begin{equation*}
    S({\bf Q},\mathrm{i}\omega_n) = \int_0^\beta e^{-\mathrm{i}\omega_n\tau}
    \langle T_\tau \hat{S}^{\phantom\dagger}_z({\bf Q},\tau)\hat{S}_z({\bf Q},0)\rangle d\tau, 
\end{equation*}
where 
\begin{equation*}
    \hat{S}^{\phantom\dagger}_z({\bf Q},\tau) = \frac{1}{\sqrt{N}}\sum_{i} 
    e^{-\mathrm{i}{\bf Q}\cdot{\bf R}_i} \left(\hat{n}_{i,\uparrow}-\hat{n}_{i,\downarrow}\right)
\end{equation*}
is the Fourier transform of the $z$-component of the local spin operator. 

The momentum-dependent charge susceptibility 
$N({\bf Q},\mathrm{i}\omega_n)$ is obtained by measuring and Fourier transforming the unequal-time 
density-density correlation function \cite{NowadnickPRL2012} 
\begin{equation*}
    N({\bf Q},\mathrm{i}\omega_n) = \frac{1}{\sqrt{N}}\sum_{i} 
    e^{-\mathrm{i}{\bf Q}\cdot{\bf R}_i}\int_0^\beta e^{-\mathrm{i}\omega_n\tau} 
    N({\bf R}_i,\tau) d\tau, 
\end{equation*}
where 
\begin{equation*}
    N({\bf R}_i,\tau) = \frac{1}{N}\sum_{j}( \langle T_\tau \hat{n}_{i+j}(\tau)\hat{n}_{j}(0)\rangle - \langle  \hat{n}_{i+j}(\tau)\rangle  \langle \hat{n}_{j}(0)\rangle).
\end{equation*}

\noindent\large\textbf{Supplementary Note 4: Additional DQMC results} \normalsize

\begin{figure}[ht]
    \centering
    \includegraphics[width=0.9\textwidth]{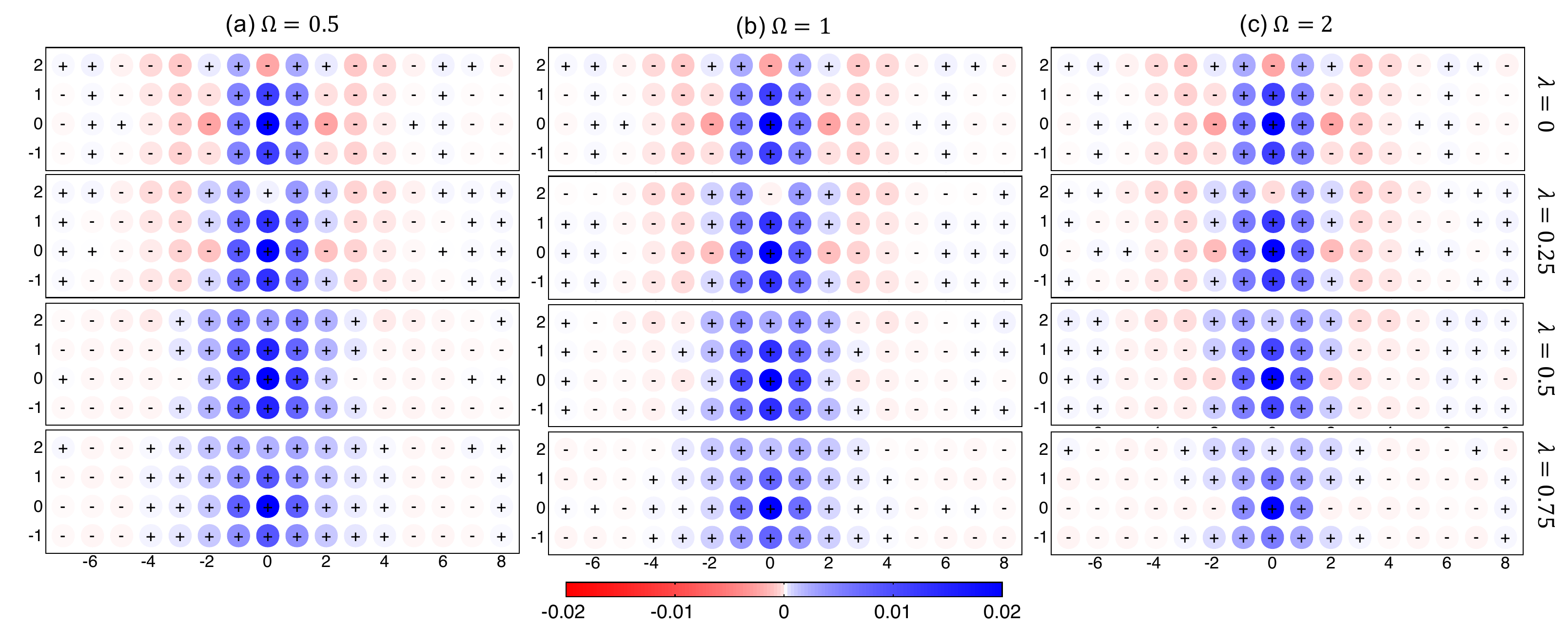}
    \caption{DQMC results for the fluctuating spin stripes in the Hubbard-Holstein model at a filling of $\langle \hat{n} \rangle = 0.8$ and an inverse temperature of $\beta = 4/t$ and obtained on $N = 16\times 4$ clusters with $t^\prime= -0.25 t$ and $U = 6t$. Each row and column shows the real-space static staggered spin-spin correlation function at different values of $\lambda$ and $\Omega$, respectively. }
    \label{fig:real_space_spin}
\end{figure}

\begin{figure}[ht]
    \centering
    \includegraphics[width=0.9\textwidth]{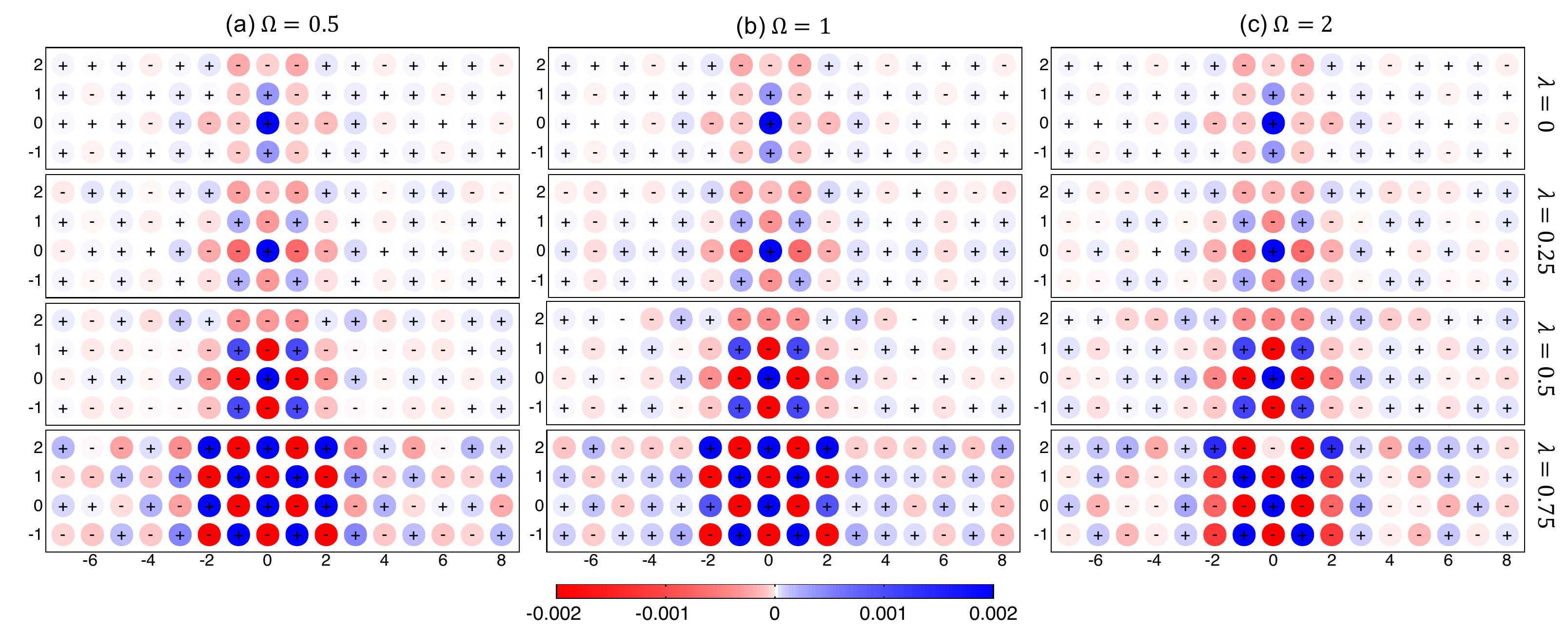}
    \caption{DQMC results for the fluctuating charge stripes in the Hubbard-Holstein model at a filling of $\langle \hat{n} \rangle = 0.8$ and an inverse temperature of $\beta = 4/t$ and obtained on $N = 16\times 4$ clusters with $t^\prime= -0.25 t$ and $U = 6t$. Each row and column show the real-space static staggered density-density correlation function at different values of $\lambda$ and $\Omega$, respectively.}
    \label{fig:real_space_charge}
\end{figure}

\begin{figure}[ht]
    \centering
    \includegraphics[width=0.5\textwidth]{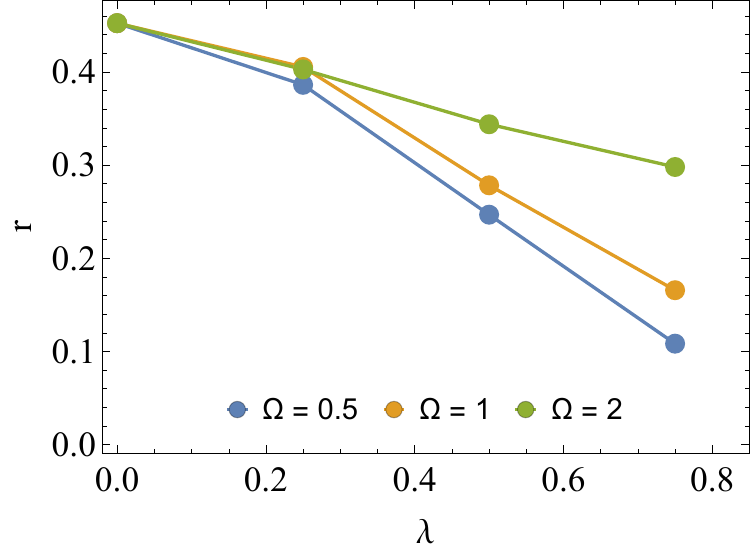}
    \caption{DQMC results of the ratio between spin and charge incommensurability $r = \delta_s/\delta_c$ in the Hubbard-Holstein model at a filling of $\langle \hat{n} \rangle = 0.8$ and an inverse temperature of $\beta = 4/t$. Results were obtained on $N = 16\times 4$ clusters with $t^\prime= -0.25 t$ and $U = 6t$ for different values of $\Omega$.}
    \label{fig:ratio}
\end{figure}

\begin{figure}[ht]
    \centering
    \includegraphics[width=0.9\textwidth]{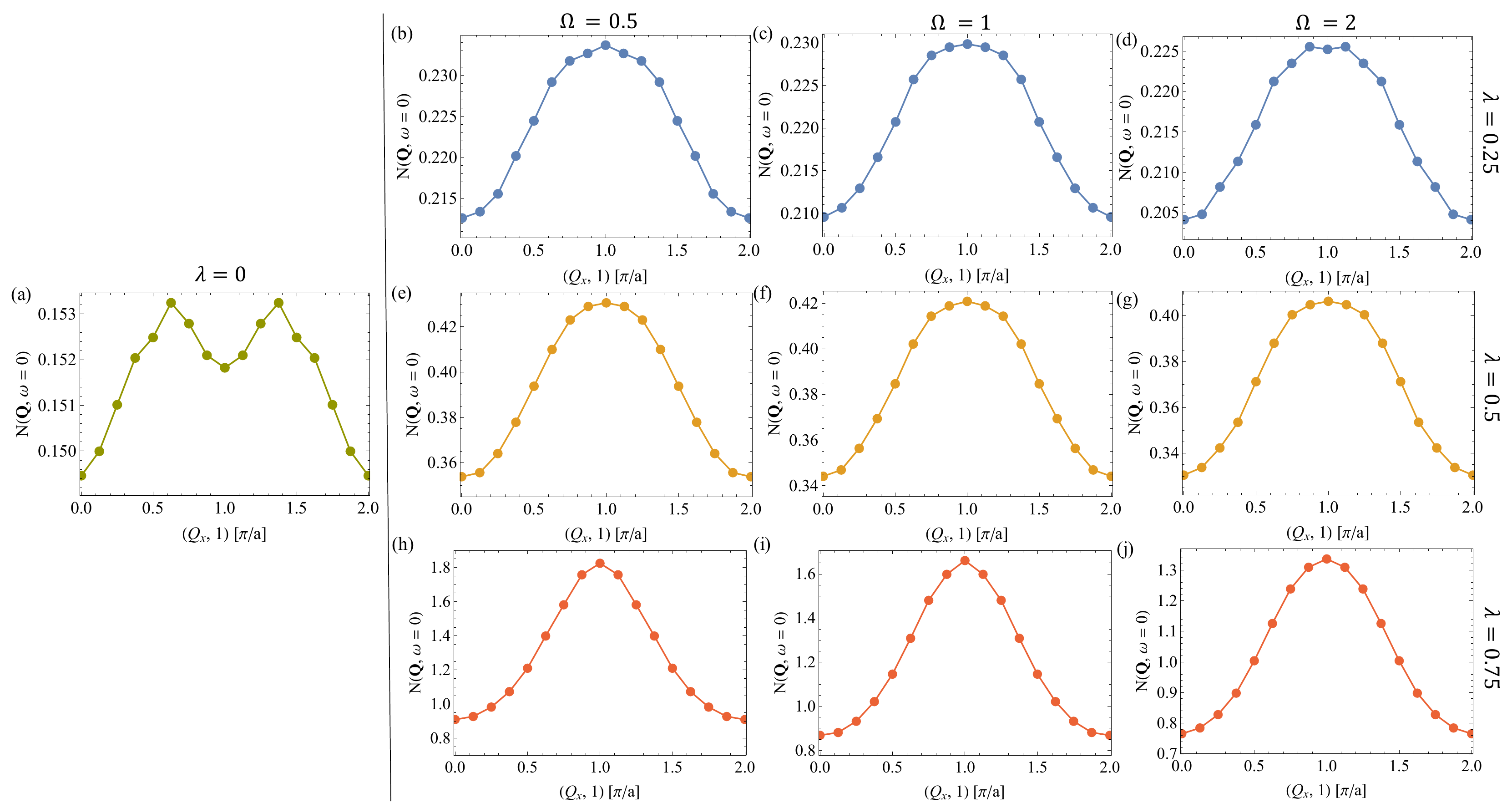}
    \caption{DQMC results for the fluctuating charge stripes in the Hubbard-Holstein model at a filling of $\langle \hat{n} \rangle = 0.8$ and an inverse temperature of $\beta = 4/t$ and obtained on $N = 16\times 4$ clusters with $t^\prime= -0.25 t$ and $U = 6t$. The panels show the static charge susceptibility at $Q_y=\pi$ as a function of $Q_x$ for different values of $\lambda$ and $\Omega$. }
    \label{fig:pipi}
\end{figure}

\clearpage
\bibliography{references}